\documentclass[amsmath,amssymb]{rsproca_new}

\usepackage{graphicx}
\usepackage{caption}
\usepackage{subcaption}
\usepackage{dcolumn}
\usepackage{bm}
\usepackage[normalem]{ulem}

\usepackage{xcolor}

\usepackage{float}

\def\half{{\textstyle \frac{1}{2}}}

\def\n{\boldsymbol{n}}
\def\m{\boldsymbol{m}}
\def\r{\boldsymbol{r}}
\def\t{\boldsymbol{t}}
\def\d{{\rm d }}
\def\e{{\rm e }}

\def\c{\boldsymbol{c}}
\def\b{\boldsymbol{b}}
\def\gth{g_{\theta\theta}}

\def\s#1{_{\textrm{#1}}}
\def\sp#1{^{\textrm{#1}}}

\def\be{\begin{equation}}
\def\ee{\end{equation}}
\def\bea{\begin{eqnarray}}
\def\eea{\end{eqnarray}}

\def\det#1{{\rm Det}( #1 )} 

\def\nl{\hfil\break}

\def\matr#1{{\ensuremath{\underline{\underline{ {\bm{#1}} }}}}}

\def\vec#1{{\ensuremath{\bm{#1}}}}

\def\dm{\matr{\delta} }
\def\g{\matr{g} } 
\def\F{\matr{F} } 
\begin{document}

\title{Inflationary routes to Gaussian curved topography}
\author{
Emmanuel Si\'efert$^{1}$ and Mark Warner$^{2}$
}

\address{
$^{1}$Lab. de Physique et M\'ecanique des Milieux H\'et\'erog\`enes, CNRS UMR7636, Ecole Sup\'erieure de Physique et Chimie Industrielles de Paris (ESPCI), Sorbonne Universit\'e, Universit\'e de Paris, 75005 Paris, France $^{2}$Cavendish Laboratory, University of Cambridge, 19 JJ Thomson Avenue, Cambridge CB3 0HE, United Kingdom
}

\subject{Mechanics, Geometry, Non-isometric origami}

\keywords{Baromorph, Curvature, Shape, Metric, Mechanics}

\corres{Mark Warner\\
\email{mw141@cam.ac.uk}}

\begin{abstract}
Abstract
\nl
Gaussian-curved shapes are obtained by inflating initially flat systems made of two superimposed strong and light thermoplastic impregnated fabric sheets heat-sealed together along a specific network of lines. The resulting  inflated structures are light and very strong because they (largely) resist deformation by the intercession of stretch. Programmed patterns of channels vary either discretely through boundaries, or continuously. The former give rise to facetted structures that are in effect non-isometric origami and which cannot unfold as in conventional folded structures,  since they present localized angle deficit or surplus.  Continuous variation of channel direction in the form of spirals is examined, giving rise to curved shells.   We solve the inverse problem consisting in finding a network of seam lines leading to a target axisymmetric shape on inflation. They too have strength from the metric changes that have been pneumatically driven, resistance to change being met with stretch and hence high forces like typical shells.
\end{abstract}


\maketitle

\section{Introduction}
Gaussian (intrinsically) curved shapes can be induced from initially flat sheets of rubber by the inflation of channels in their interior by air that selectively distorts their neighbouring environment. These have been termed ``baromorphs" \cite{Siefert_baromorphs2019}. Air pumps thus offer a rapid, simple and reversible way of creating non-trivial changes in metric of such a space, and hence routes to shapes that would otherwise be inaccessible from flat space.

Another, related method, which is simpler and which creates very light and strong analogues of baromorphs, is to take two sheets of thermoplasticurethane (TPU)- impregnated nylon fabric that is air-tight. Placed one on top of the other, these can be welded together  using a soldering iron mounted on an XY-plotter to create channels in practically any desired pattern that can then be inflated. Isolated channels can be shape-programmed, that is creating a desired 1-D curvilinear shape on inflation \cite{Siefert_PNAS2019}. Putting patterns of channels together gives associated contractions on inflation that cause a (coarse-grained) metric change and hence shape evolution which we explore here and which are addressed in \cite{Siefert_backpacks2019}.

Other routes to differential in-plane distortion and hence to intrinsically-curved topography have been examined. Some are non-reversible, such as in the growth of leaves that wrinkle because their azimuthal growth exceeds their radial growth, leading to a circumference too long relative to their radii \cite{dervaux2008morphogenesis}. Other reversible examples include the differential swelling of hydrogels \cite{klein2007shaping,kim2012designing},  the controlled in-plane expansion of dielectric elastomers through spatially varying electric fields \cite{bense17,Hajiesmaili_2019}, and the response of patterned liquid crystal elastomers (LCE) sheets where heat or light give length changes \cite{warner2007liquid,Finkphoto} which, when differential, give bend \cite{vanOosten:07} or topographical \cite{modes2010disclinations,de2012engineering,Ahn2016photo} deformations. The hydrogel distortions are isotropic (but see \cite{Gladman2016}), with the extent of volume change spatially varying due to varying linkage density.  Liquid crystal elastomer (LCE) or liquid crystal glass (LCG) systems typically have the same magnitude of anisotropic distortion throughout (the same temperature or illumination throughout), but the direction of anisotropy  can vary spatially  in-plane.

Here we further explore pneumatic ideas and methods as in \cite{Siefert_backpacks2019}, making contact with analogous work on LCEs. We give routes to topographical changes due to:
\begin{enumerate}
  \item Discrete changes of direction between regions of uniform, anisotropic in-plane distortion cause folds along lines when activated \cite{modes2011blueprinting} --  in effect ``non-isometric origami" \cite{Plucinsky:16} since the faces change dimensions and resist with stretch at any attempt to unbend.
  \item Continuous, curved channels that produce a distortion constant in amplitude but varying in direction.
\end{enumerate}
The origami is of a completely different character from conventional, isometric folding of a medium (typically paper) at constant length. It is also different from modern, activated-fold origami\cite{miskin2018graphene} which is also weak since it relies on bend (in the folds that act as hinges) rather than stretch. For non-isometric origami, see the reviews \cite{modes2016review,Warner2019review}.  The Gaussian Curvature (GC) is concentrated at vertices where there are angular deficits.

The continuous field distortions give, in general, shells with continuous GC distributions and stiffnesses of structures due to stretches arising from distorting a curved shell rather than from bend, which is weak. We will see that one can track the evolution of curves to form geodesics, and other phenomena associated with space curvature changes.

We address both the forward problem of what the shape is, given a channel distribution, and the inverse problem of what channel distribution is required to yield a given shape on inflation.

\section{Materials, deformations, a simple metric change}\label{sect:materials}
\begin{figure}
\centering
\includegraphics[width=0.85\linewidth]{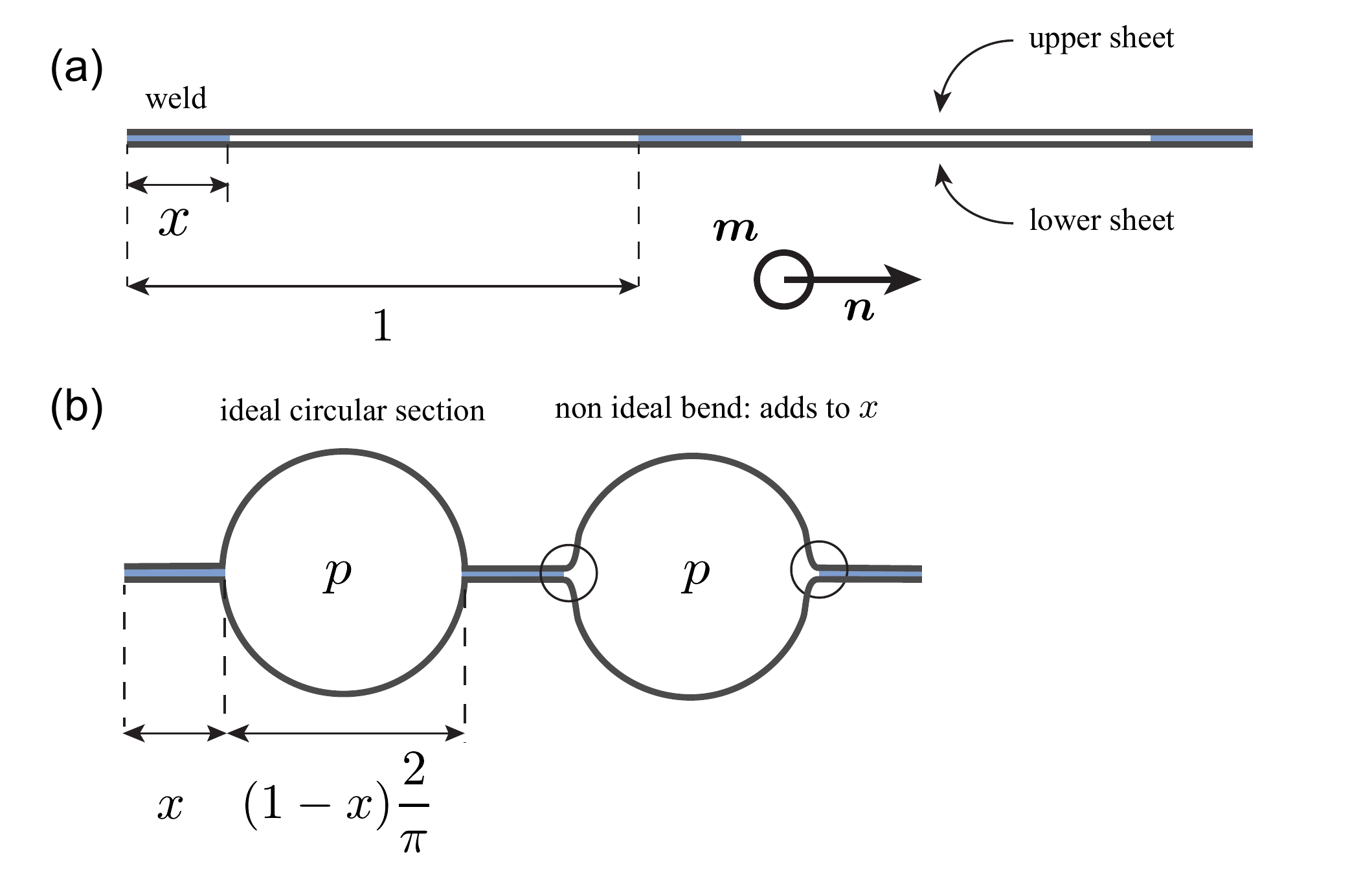}
  \caption{Schematic principle of the deformation upon inflation: the unit repeat length has a fraction $x$ of weld and $1-x$ of free sheet. (a) A section (channels into the paper) before inflation; (b) a section after inflation. The direction of contraction, $\n$ is equivalent to a nematic director in an LC solid. The contraction of the $1-x$ fraction is by a factor of $2/\pi$ in the ideal case, or by a factor $ > 2/\pi$ if there is length taken up by bend and the welded fraction remains unchanged. In the direction $\m$ of the seam lines, no contraction occurs.}
  \label{fig:set_up}
\end{figure}

Figure \ref{fig:set_up} shows quasi-inextensible heat-sealable sheets in section, a fraction $x$ dedicated to the weld and $1-x$ being free for inflation.
The resultant circular section -maximising the volume- of a channel has a circumference, assuming quasi-inextensibility, of $2(1-x)$ and hence a diameter a factor of $\pi$ smaller.  Here, we neglect the slight influence of the curvature of the seam lines on the cross-section profile and hence on the coarse-grained contraction, investigated in \cite{Siefert_PNAS2019}.The contraction factor in the direction shown, perpendicular to the channels and denoted by $\n$, is accordingly:
\be
\lambda = (1-x) 2/\pi + x \rightarrow 2/\pi \label{eq:contraction}
\ee
where the limit is for narrow welds, $x \rightarrow 0$. The fabric is essentially inextensible and accordingly $\lambda = 1$ for the other in-plane direction (into the paper) \cite{Siefert_backpacks2019}. The in-plane, $2\times 2$ deformation gradient $\F$ and the associated metric tensor $\g = \F^T\cdot \F$ are:
\bea
\F&=& (\lambda-1)\n\otimes\n+\dm \label{eq:deformation}\\
\g &=& (\lambda^2-1)\n\otimes\n+\dm \label{eq:metric},
\eea
where $\dm$ denotes the identity operator on $\mathbb{R}^2$. The metric describes changed lengths and, along with its derivatives, the GC that develops \cite{modes2012responsive,aharoni2014geometry,Mostajeran2015}.

\begin{figure}
\centering
\includegraphics[width=1\linewidth]{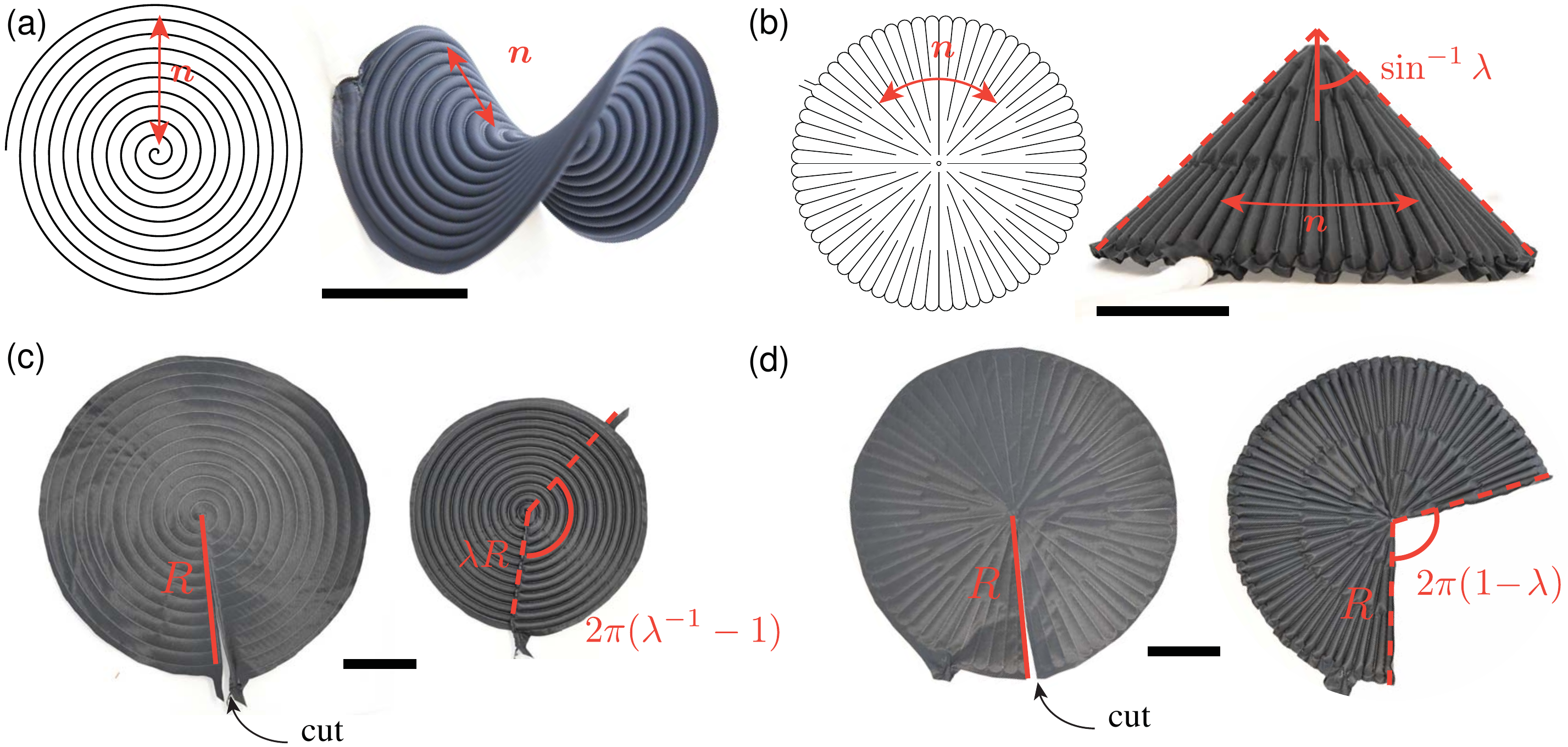}
  \caption{Radial and azimuthal directors: (a) An Archimedean spiral approximating azimuthal channels (radial $\n$), results  in  an anticone (``ruff") on inflation, (b) Radial channels (azimuthal $\n$) induce a cone upon inflation of half summit angle $\varphi= \sin^{-1}\lambda$.
  (c) Angular surplus or (d) angular deficit, made apparent when discs with the same respective patterns have instead a radial cut and do not change their topography. Structures are made of TPU-impregnated nylon fabrics (40den, 70g/sqm from Extremtextil). Scale bars: 5 cm. }
  \label{fig:cones}
\end{figure}

The simplest examples of pneumatic GC generation are those of (a) azimuthal, and (b) radial channels (radial and azimuthal directors respectively). Figure~\ref{fig:cones} shows realisations of these examples, with $\n$ vectors superimposed, before and after inflation.
In (a), the (ideal) radial contraction of $r \rightarrow \lambda r = (2/\pi) r$ leaves the circumference, $2\pi r$, longer than required and an anticone (a') results, that is, a surface with localised negative GC analogously to in LCEs \cite{modes2010disclinations,PhysRevE.92.010401}. There is an angular surplus and hence (negative) GC of magnitude $\pi^2 - 2\pi$, see Fig.~\ref{fig:cones} (a).

In (b), a circumference $2\pi r \rightarrow (2/\pi) \cdot 2\pi r = 4r$, with an unchanged radius $r$ giving a circumferential deficit. The angular deficit and hence (positive) GC is $2\pi - 4$. In the uncut case, the cone formed has semi angle $\phi = \sin^{-1}(2/\pi)$, since the in-material radius remains $r$, but the circumference is generated by an in-space radius $(2/\pi) r$, the ratio of the two radii giving $\sin\phi$.

From these simplest examples of topography change, we can already see mechanics emerging. The cones, with their radial channels, are relatively easy to deform; their circular sections change under transverse forces since the welds between the channels can easily bend. Loading from the tip is resisted by circumferences resisting extension which would result in GC change and hence circumferential stretch (which in turn is resisted pneumatically). Equally, the anti-cone case of azimuthal channels is rigid because, despite there being welds that would bend easily, any such bend would alter the GC and induce in this case compressions along the channels. These too are resisted pneumatically since volume too would thereby be diminished. One can show that spiral patterns combine the best of both of these scenarios.

\section{Non-isometric origami}\label{sect:origami}
\begin{figure}
\centering
\includegraphics[width=0.75\linewidth]{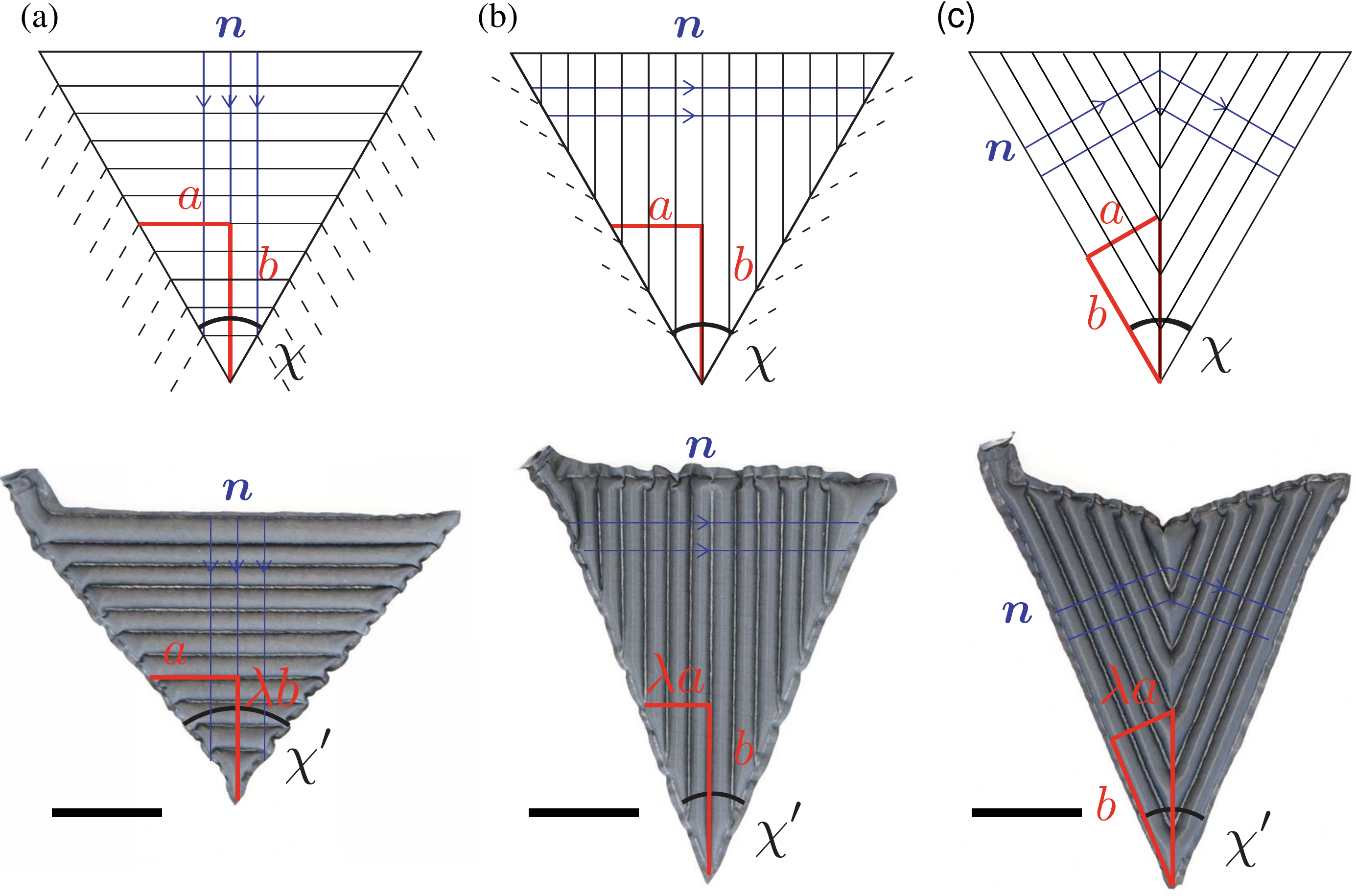}
  \caption{Sectors of uniform director, simply arranged parallel (a) or perpendicular (b) to the bisector, or parallel to the sides (c), necessarily having an internal line of R-1C along the bisector. The director fields just outside the sectors are shown, indicating how these regions connect to their neighbours across the boundary vectors $\t$. The right triangles with sides $a$ and $b$ give simple rules for the change of apex semi angle $\chi/2$ to $\chi'/2$ after distortion. Below each is a pneumatic realisation, with triangular sections opening out (a), or closing in (b) and (c), with the last example having the upper side of the triangle developing an angle. Structures are made of TPU-impregnated nylon fabrics (40den, 70g/sqm from Extremtextil). Scale bars: 2 cm}
  \label{fig:vocab}
\end{figure}
Piecewise uniform channel/director fields offer (localised) GC at vertices where the fields meet \cite{modes2011blueprinting,Plucinsky:16,Plucinsky:17,Plucinsky:18}. Compatibility between the regions on distortion requires Rank-1 Connectedness (R-1C) at the boundaries where the uniform fields intersect. Physically, this R-1C condition is that deformations parallel to the boundary must be identical from one side (1) to the other (2) to prevent tearing. It is expressed, in this nematic analogue, as $\n_1\cdot \t = \n_2\cdot \t$, where $\t$ is the vector along the line of separation: The directors must make equal angles with the boundary.

Simple units must be put together in a R-1C way. Three such are shown in figure~\ref{fig:vocab} where the change in angle is trivial to calculate \cite{modes2011blueprinting} from the right triangles shown where opposite or adjacent sides change by a factor $\lambda$, giving the change in $\tan\chi/2$.

Hence in (a) and (c) (which are essentially equivalent) one has $\chi' = 2\tan^{-1}\left(\frac{1}{\lambda}\tan(\chi/2)\right)$ and for (b) one has $\chi' = 2\tan^{-1}\left(\lambda\tan(\chi/2)\right)$, with $\lambda = 2/\pi$ in the ideal case. Several examples are given in \cite{modes2011blueprinting} of how elementary units can fit together to give flat sheets that transform into complex shapes with vertices.

\begin{figure}
\centering
\includegraphics[width=\linewidth]{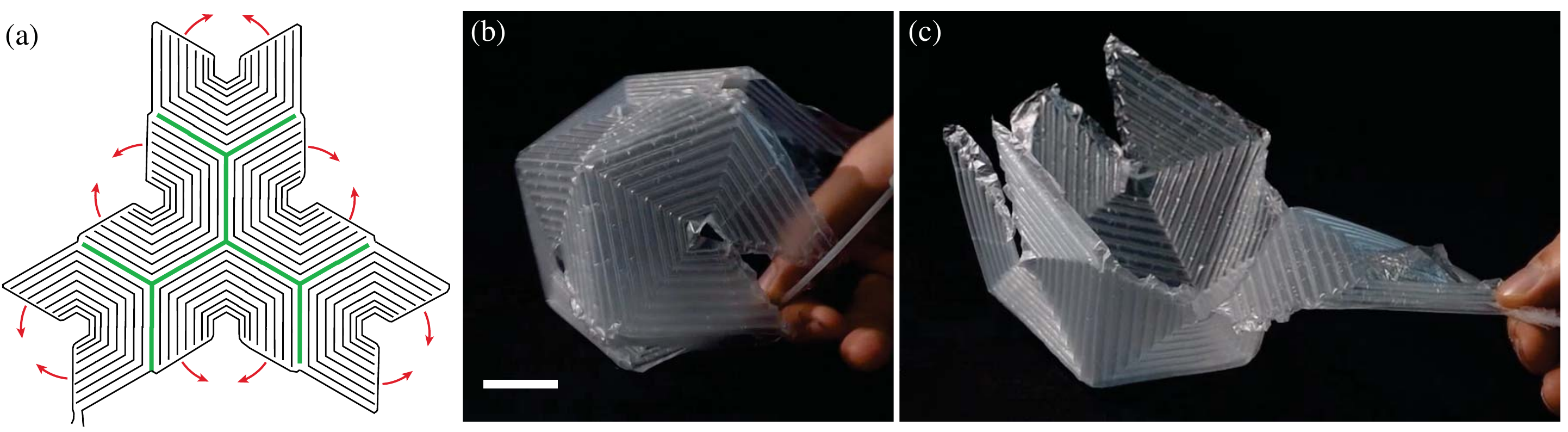}
  \caption{Attempting a non-isometric origami cube. In contrast with LCE, the inflated structure tends to bend along the weak lines of the seam and not at the R-1C boundary. (a) Pattern for a non isometric cube, the green lines highlighting the position of the edges. (b) As expected, the contraction ratio is not large enough to close the structure made of 4 $\mu m$ thick polyethylene sheets. (c) Every vertex may be inverted since no bending direction is preferred (from the symmetry of the structure across the thickness). Scale bar: 2cm. }
  \label{fig:cube}
\end{figure}

\begin{figure}
\centering
\includegraphics[width=0.9\linewidth]{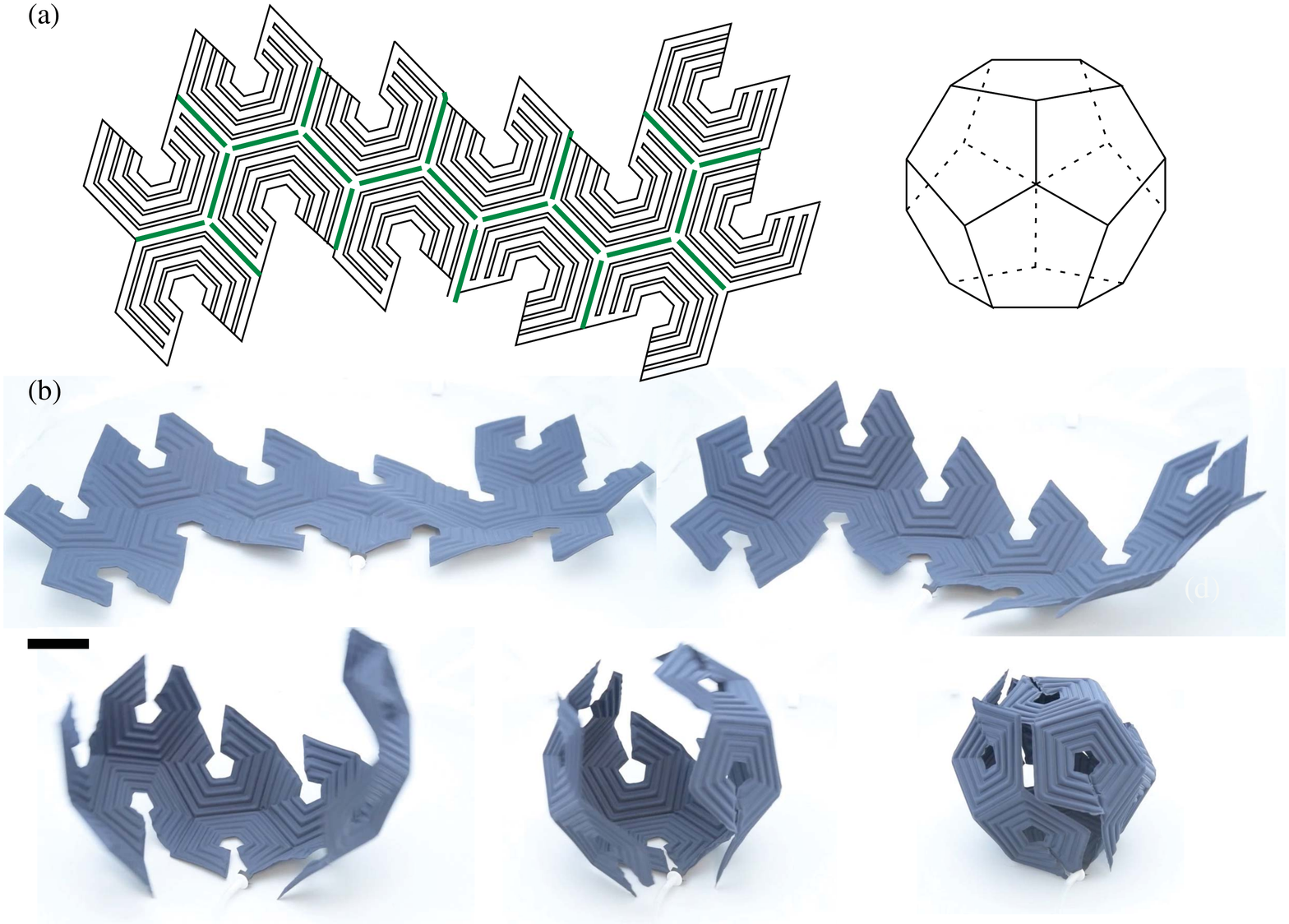}
  \caption{(a) A pattern of channels schematically in flat space (left), that will inflate to a dodecahedron (right). The green lines show the position of the edges. (b) Upon inflation, the structure gradually closes to form the target dodecahedron. Structures are made of TPU-impregnated nylon fabrics (70den, 170g/sqm from Extremtextil). Scale bar: 5 cm}
  \label{fig:dodecahedron}
\end{figure}

A cube has three square faces meeting at each corner, so each sector has to develop a $\chi' = \pi/2$. The three proto-faces meeting at a point while the sheet is still flat must have angles $\chi = 2\pi/3$. Using the above transformation of the type (b), one requires a $\lambda = \tan(\chi'/2)/\tan(\chi/2) = \tan(\pi/4)/\tan(\pi/3) = 1/\sqrt{3} = .577$ which is less than the available $2/\pi = 0.636$ -- cubes are inaccessible to backpack material inflatables; see Fig.~\ref{fig:cube}, where also the question of up-down or in-out deformation degeneracy arises.
Dodecahedra have pentagonal faces, three of which meet at a vertex. There being three, in the flat state $\chi = 2\pi/3$. In the distorted state, the pentagonal angle is $\chi' = 3\pi / 5$. The required $\lambda = \tan(3\pi/10)/\tan(\pi/3) = 0.794$  is easily accessible, as is demonstrated in figure~\ref{fig:dodecahedron}.

\begin{figure}
\centering
\includegraphics[width=0.9\linewidth]{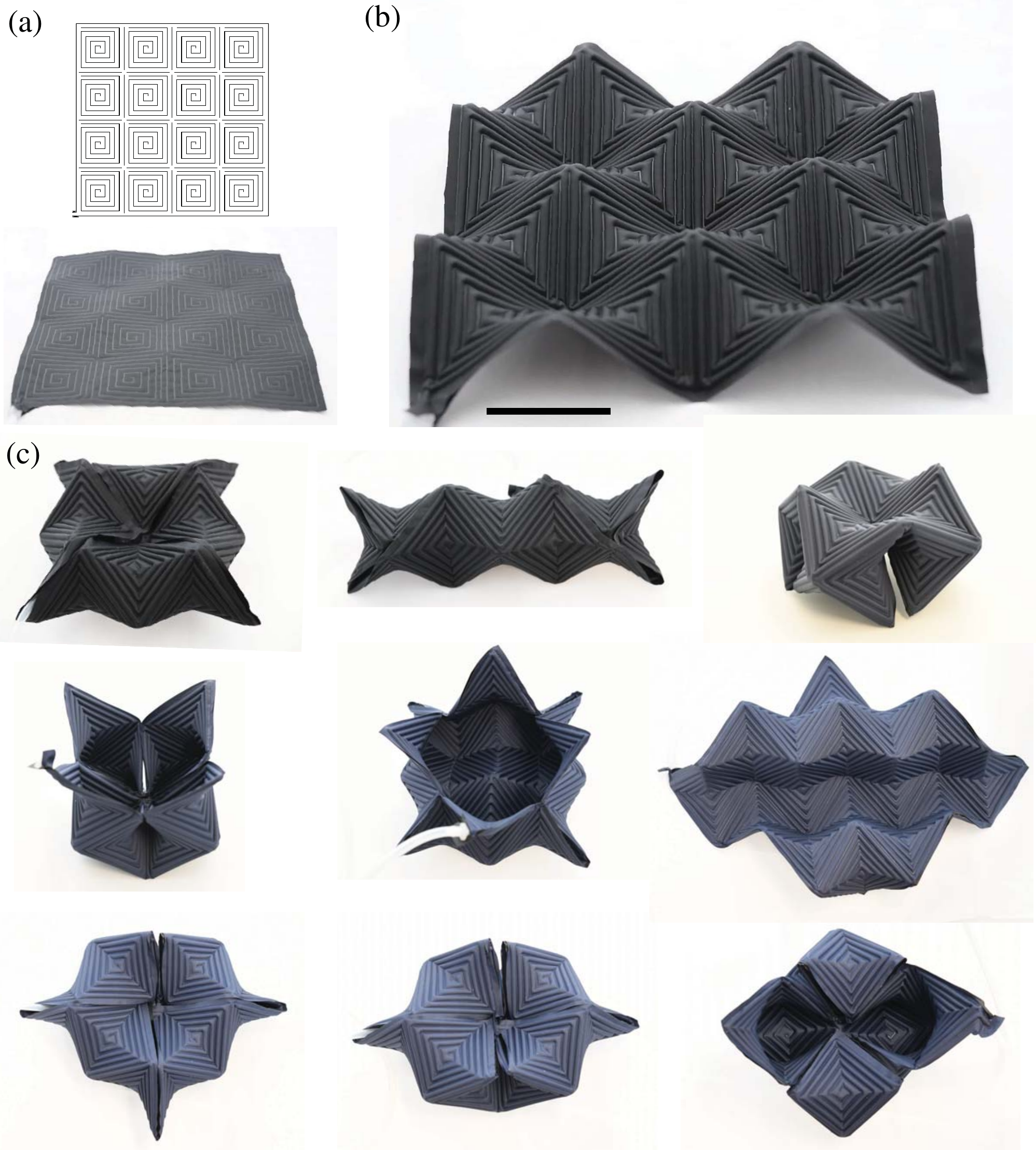}
  \caption{An array of alternating positive ($+1$) and negative ($-1$) topological defects made of TPU-impregnated nylon fabric sheets (70 den, 170g/sqm from Extremtextil). (a) Pattern and picture of the flat state. (b) Upon inflation, the structure exhibits mountain peaks and troughs ($+1$) separated by saddles ($-1$)  (See Supplementary Video eggbox\_top\_view.MP4). Scale bar: 5cm. (c) Each saddle may be snapped to invert the two upper and lower corners and the structure may fold along the seams connecting peaks and troughs enlarging remarkably the family of stable shapes that may be reached.}
  \label{fig:eggs}
\end{figure}

Arrays of R-1C sectors can also form extended topographies, see figure~\ref{fig:eggs}(a) and (b),  rather than closed structures such as in figures \ref{fig:cube} and \ref{fig:dodecahedron}.
Inspecting the array of channels, or their orthogonal dual -- the director, one sees an array of $\pm 1$ topological defects in 2-D. In the inflated egg crate array, one sees these correspond to mountain peaks and troughs ($+1$) separated by saddles ($-1$). The rigidity of these units (especially the saddles) is further enhanced by their being in an array. Attempts to deform the system induce stretch, which is very strong. Such systems are reminiscent of the LC solid arrays of White \textit{et al} \cite{Ware2015,guin2018_lifters} which can lift thousands of times their own weight when loaded and actuated\cite{guin2018_lifters} . See also the 3-D printed LC actuator arrays of Kotikan \textit{et al} \cite{Kotikan_2018}.
 In addition to making an egg crate, each saddle may be snapped to invert the two upper and lower opposite corners and the structure may also fold along the seams connecting peaks and troughs enlarging remarkably the family of stable shapes that may be reached. In order to go from one shape to the other, at least one saddle should be snapped, which means that each shape corresponds to a local minimum in the energy landscape. A few examples of shapes are shown in Fig. \ref{fig:eggs}(c). There are some constraints in this system and the snapping of one saddle may impose the snapping of its neighbours. The count of all the possible stable states has not been carried out yet and will be the subject of future work.

\subsection{Non-regular vertices and the rules they obey}
\begin{figure}
\centering
\includegraphics[width=0.8\linewidth]{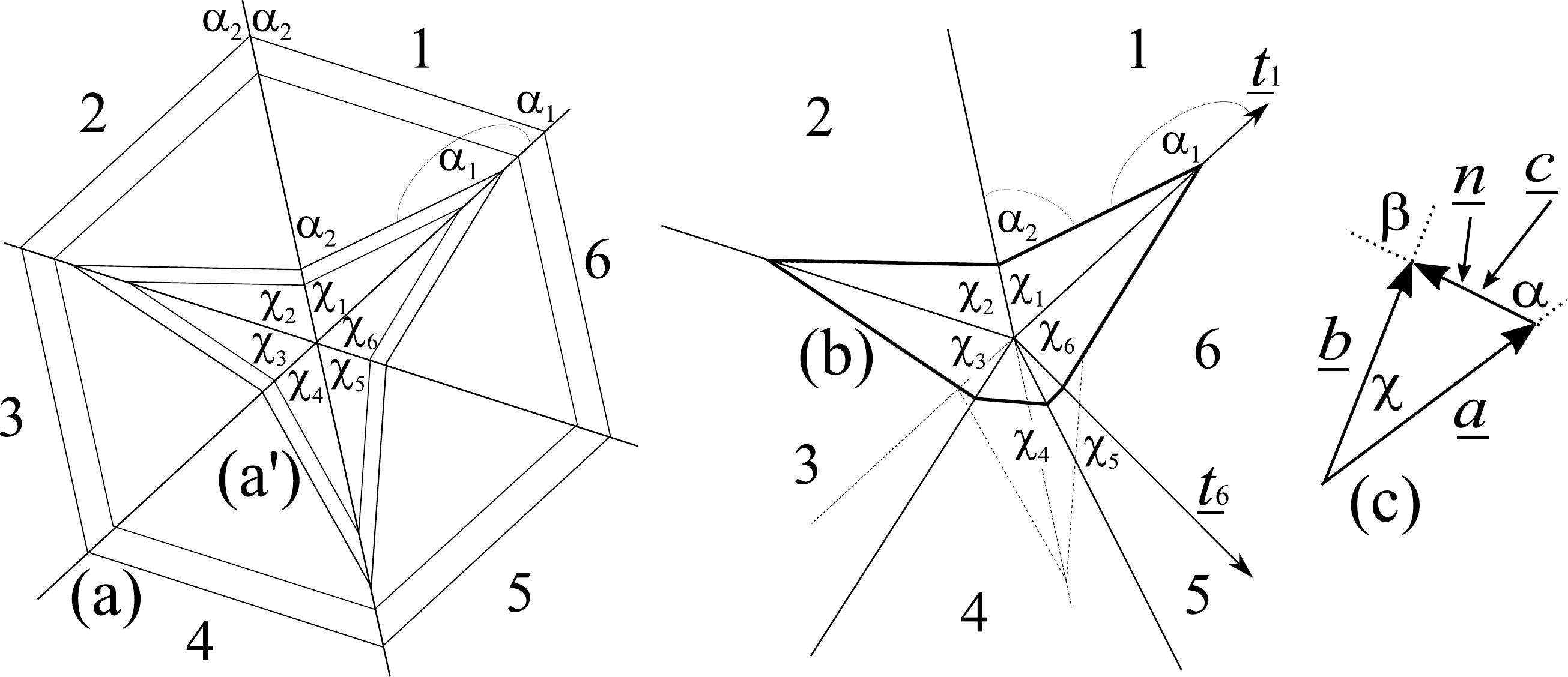}
  \caption{(a) A regular proto-vertex (even, $n=6$, all internal sector angles $\chi$ identical) with a regular director field (all $\alpha$s equal), and (a') a star-shaped field where there are two $\alpha$s. The angles $\alpha_i$ are those made by the director with the divider between sectors $(i-1)$ and $i$.  (b) An irregular proto-vertex  (even, $n=6$) with internal angles $\chi_i$ satisfying rules given in the text, and with corresponding director angles $\alpha_i$ set by the same rules. The vectors $\t_i$ define the boundaries. The dotted lines are vestiges from the regular case (a) that we have deviated from. (c) A sector triangle from which to calculate $\chi \rightarrow \chi'$. The opposite side, $\c$, is parallel to $\n$.}
  \label{fig:general}
\end{figure}

Generally, vertices do not have to be formed by the meeting of identically-shaped regions of uniform director. But deviation from regularity can only be within strong conditions on the internal angles of sectors, $\chi_i$ and on the angles $\alpha_i$ that the director $\n_i$ in the $i\sp{th}$ sector makes with the interface between the $i-1\sp{th}$ and $i\sp{th}$ sectors, that is conditions on $\n_i \cdot \t_i$; see figure~\ref{fig:general}.
The director angles $\alpha_{i+1}$ and $\alpha_i$ are connected by
\bea
\alpha_{i+1}&=& \pi + \chi_i - \alpha_i \label{eq:relation}\\
\textrm{even} \; i : \;\; \alpha_{i+1}&=& \chi_i - \chi_{i-1} + \dots +\chi_2 - \chi_1 + \alpha_1 \label{eq:even}\\
\textrm{odd} \; i : \;\; \alpha_{i+1}&=& \pi + \chi_i - \chi_{i-1} + \dots -\chi_2 + \chi_1 - \alpha_1 \label{eq:odd}.
\eea
The first expression arises from considering the triangle formed in the $i\sp{th}$ sector by the two boundaries and the director line, the latter two from  iteration for $i$ even or odd respectively.

For an $n$-sector vertex, $\alpha_{n+1} \equiv \alpha_1$, whereupon the $n$ even and odd cases yield:
\bea
\textrm{even}\; n: \;\;\;\; 0 &=& \chi_n - \chi_{n-1} + \dots +\chi_2 - \chi_1  \label{eq:even_condition}\\
\textrm{odd} \; n: \;\; \alpha_1&=& \pi/2  + \half\left( \chi_n - \chi_{n-1} + \dots -\chi_2 + \chi_1\right), \label{eq:odd_condition}
\eea
with of course the condition $\Sigma_{i = 1}^n \chi_i = 2\pi$. Thus for vertices with even number of sectors, $\alpha_1$ can take any value, and the $\alpha_{i\ne 1}$ then follow from eqns~(\ref{eq:even}) and (\ref{eq:odd}). For odd cases, the choice of  $\alpha_1$ is constrained by condition~(\ref{eq:odd_condition}) and the subsequent $\alpha_{i\ne 1}$ again follow from eqns~(\ref{eq:even}) and (\ref{eq:odd}).

 Figure~\ref{fig:general}(a) shows an even, regular case where $n=6$ and $\chi = \pi/3$ in each sector. Choosing $\alpha_1 = 2\pi/3$ gives all sectors the same $\alpha$, or choosing $\alpha_1 > 2\pi/3$ gives $\alpha_2 < 2\pi/3$, with subsequent alternation. By contrast, (b) has $\chi_1 = \chi_2 =\chi_4 = \pi/3$ as before, but $\chi_3$, $\chi_2$ and $\chi_6$ changed from $\pi/3$ such that condition (\ref{eq:even_condition}) is respected, whence the director lines close and are R-1C as required.

 To calculate the GC associated with a vertex, one must calculate each $\chi_i'$ and hence the total angular deficit/surplus $\Sigma_{i=1}^n\chi'_i - 2\pi$. Not all the sectors' $\n$ fields are of the symmetric form of Fig.~\ref{fig:vocab}, and one requires the metric tensor in its non-diagonal form. Considering the triangle of Fig.~\ref{fig:general}(c), the side $\c$ is parallel to $\n$ and hence simply changes its length as $c \rightarrow \lambda c$. The other sides change as, for instance, $b^2 \rightarrow \b \cdot \g \cdot \b = f_b^2 b^2$ with $f_b^2 = \lambda ^2 \cos^2\beta + \sin^2\beta$ if one considers the factors of $\b\cdot \n$ required in contracting with the metric tensor, eqn.~(\ref{eq:metric}), with $\beta=\alpha-\chi$. Applying the cosine rule in the triangle before and after deformation then gives:
 \be
 \cos\chi' = \frac{\lambda^2}{f_a f_b}\cos\chi + \frac{(f_a^2 - \lambda^2  )a^2 + ( f_b^2-\lambda^2 )b^2 }{2ab} \label{eq:chi_new} .
 \ee
 The second term depends on the ratio $a/b$ and its inverse, so  the actual size of the sector is irrelevant.

 Experimental realizations of irregular vertices have not been conducted yet and will be the subject of future work. Irregular vertices indeed strongly enrich the possibilities of non-isometric origami.

 \section{Spiral patterns of channels/directors}\label{sect:spirals}
 Spiral channels offer great advantages since (a) for a circularly symmetric pattern, one can vary the angle the director makes with the radial direction, $\alpha(r)$, and thereby create complex GC distributions, and (b) with channels varying in direction, there are not weak directions where bend can occur simply along straight seams.

\begin{figure}
\centering
\includegraphics[width=0.6\linewidth]{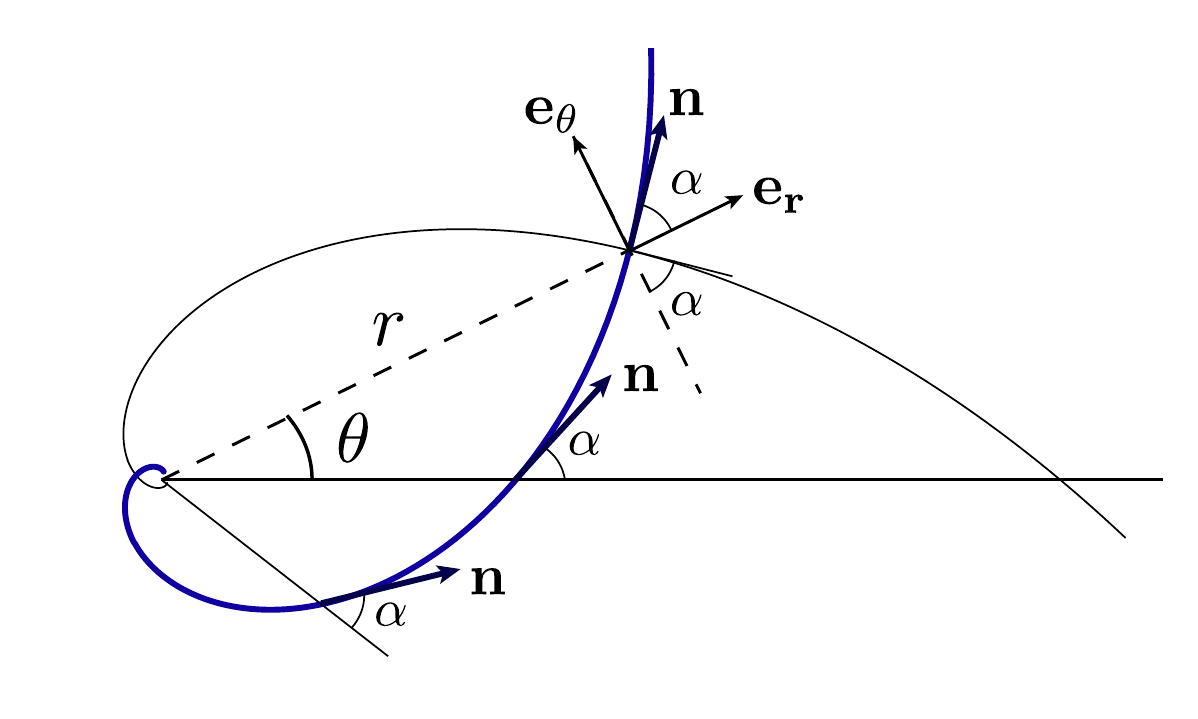}
  \caption{A logarithmic spiral director $\n$ (heavy, blue on-line) forming a constant angle $\alpha$ with the radial vector $\bf e_r$. The corresponding seam (black, light line) is a clockwise turning logarithmic spiral forming an angle $\pi/2 -\alpha$ with $\bf e_r$, that is, the orthogonal dual to $\n$.
 }
  \label{figlog}
\end{figure}

 \subsection{Logarithmic spirals}\label{subsect:log-spirals}
  Logarithmic spirals are a simple extension of the illustrative circular and radial director patterns ($\alpha = \pi/2, 0$ respectively) in section \ref{sect:materials}. They have an $\alpha =$ const. ($\ne 0, \pi/2$) that is not a function of reference space radial distance $r$. Hence the ratio, after inflation, of in-material (intrinsic) radius and circumference is also constant with $r$, and  they evolve into cones or anti-cones, depending on the value of $\alpha$. This evolution is more complex than before. In the reference state frame of the radius, $\hat{\r}$, and azimuth, $\hat{\vec{\theta}}$, the metric tensor now has off-diagonal elements. There is a differential (with $r$) rotation of material points as one inflates. Final state, in-material radii evolve from proto-radii that are (non-geodesic) curves in flat space. Tangents $\t$ to curves in the flat reference state evolve as $\F\cdot\t$, where the tangent to circles in polar coordinates $(r,\theta)$ are $\t_c = (0,1)$, and general curves have $\t = ( \d r(s)/\d s, \d\theta(s)/\d s)$, where $\t$ is a unit vector since we have taken a unit speed parametrisation in terms of arc length $s$. Since the director pattern is circularly symmetric, circles evolve to inflated/deflated circles.

 For a given curve to evolve to an in-material radius, its evolved tangent must be orthogonal to that ($\t_c$) of circles, that is $(\F \cdot \t_c )^T\cdot(\F\cdot \t)  \equiv \t_c^T \cdot \g \cdot \t = 0$; see the discussion in \cite{Mostajeran2016,Mostajeran2018}. The tangents $\t$ and $\t_c$ are orthogonal under the metric $\g$.
 In polar coordinates, $\g$ has elements:
 \begin{align} \label{polarMetric}
g_{rr}&=\lambda^2 \cos^2\alpha +\sin^2\alpha, \nonumber \\
g_{r\theta} =g_{\theta r}&= - \frac{r}{2}\left(1-\lambda^2\right)\sin2\alpha, \nonumber \\
g_{\theta\theta}&=r^2\left[ \lambda^2\sin^2\alpha + \cos^2\alpha\right].
\end{align}
 The condition above for $\r(s)$ to be the ancestor of a radial geodesic is then \cite{Mostajeran2016}
 \bea
g_{\theta r} \d r(s)/\d s  +  g_{\theta\theta}\d\theta(s)/\d s& =& 0\nonumber\\
\rightarrow \d r/d\theta &=& -   g_{\theta\theta}/g_{\theta r} =  2r \frac{\lambda^2\sin^2\alpha + \cos^2\alpha}{ \left(1-\lambda^2\right)\sin2\alpha} .\label{eq:geodesic}
\eea
The expression~(\ref{eq:geodesic}) is for a general $\alpha(r)$ but is particularly simple when the director integral curves are log spirals. Then $\alpha$ is constant and the director follows
\be
r(\theta) = r(0) \e^{b \theta}, \;\; \textrm{with} \;\; b = \cot(\alpha) \label{eq:integral_log_spiral}.
\ee
Then integrating eqn.~(\ref{eq:geodesic}) gives the equation of proto radius in the initial state:
\be
r(\theta) = r(0) \e^{c \theta}, \;\; \textrm{with} \;\; c = \frac{\lambda^2 + b^2}{b(1-\lambda^2)}, \label{eq:proto_log_spiral}
\ee
also a log spiral, but with $c \ne b$, where $c = \cot\beta$ defines the angle $\beta$ of the proto-radius log spiral.

\begin{figure}
\centering
\includegraphics[width=\linewidth]{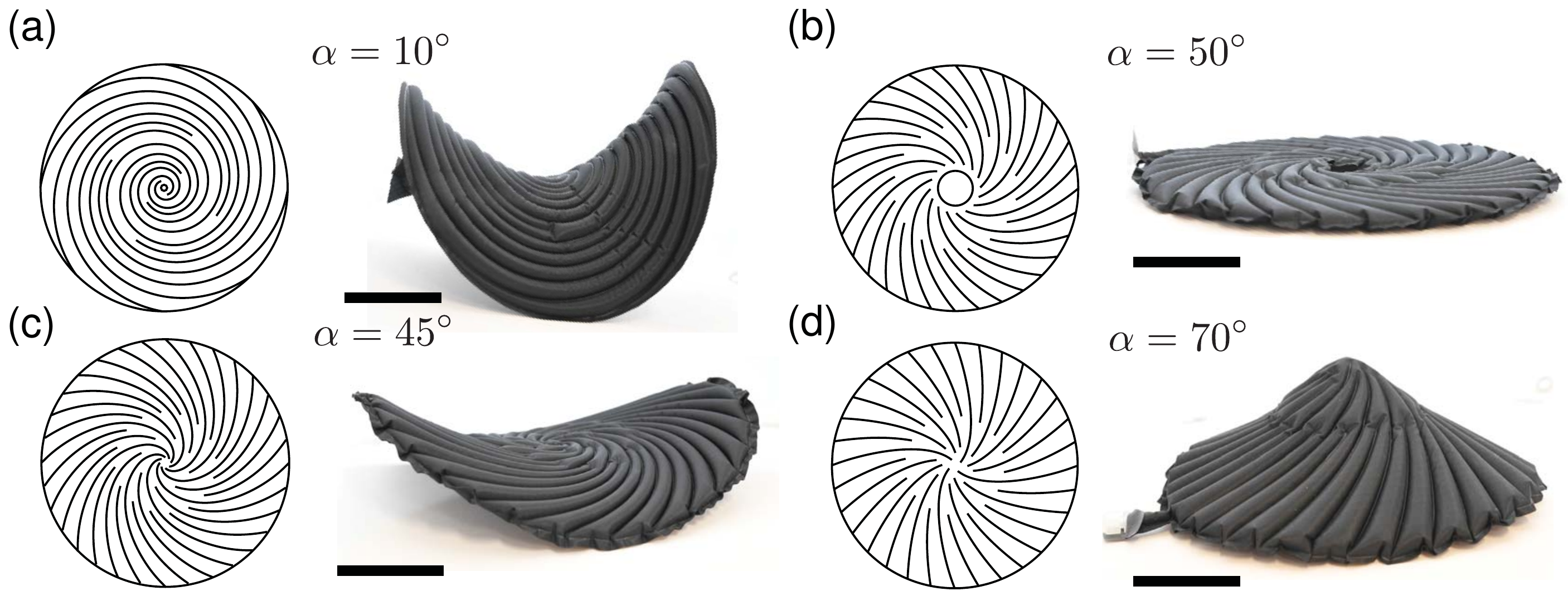}
  \caption{Anti-cones and cones arising from channel systems with log spiral directors with (constant) angles $\alpha$ to the radii shown. Note that the channels themselves are at the complement angle of $\alpha$. The critical angle where the inflated structure remains flat is an $\alpha\s{c} \approx 50^{\circ}$. See the extreme cases of director angle in Fig~\ref{fig:cones}. Structures are made of TPU-impregnated nylon fabrics (40den, 70g/sqm from Extremtextil). Scale bars: 5 cm.
 }
  \label{fig:log-spirals}
\end{figure}

\begin{figure}
\centering
\includegraphics[width=0.8\linewidth]{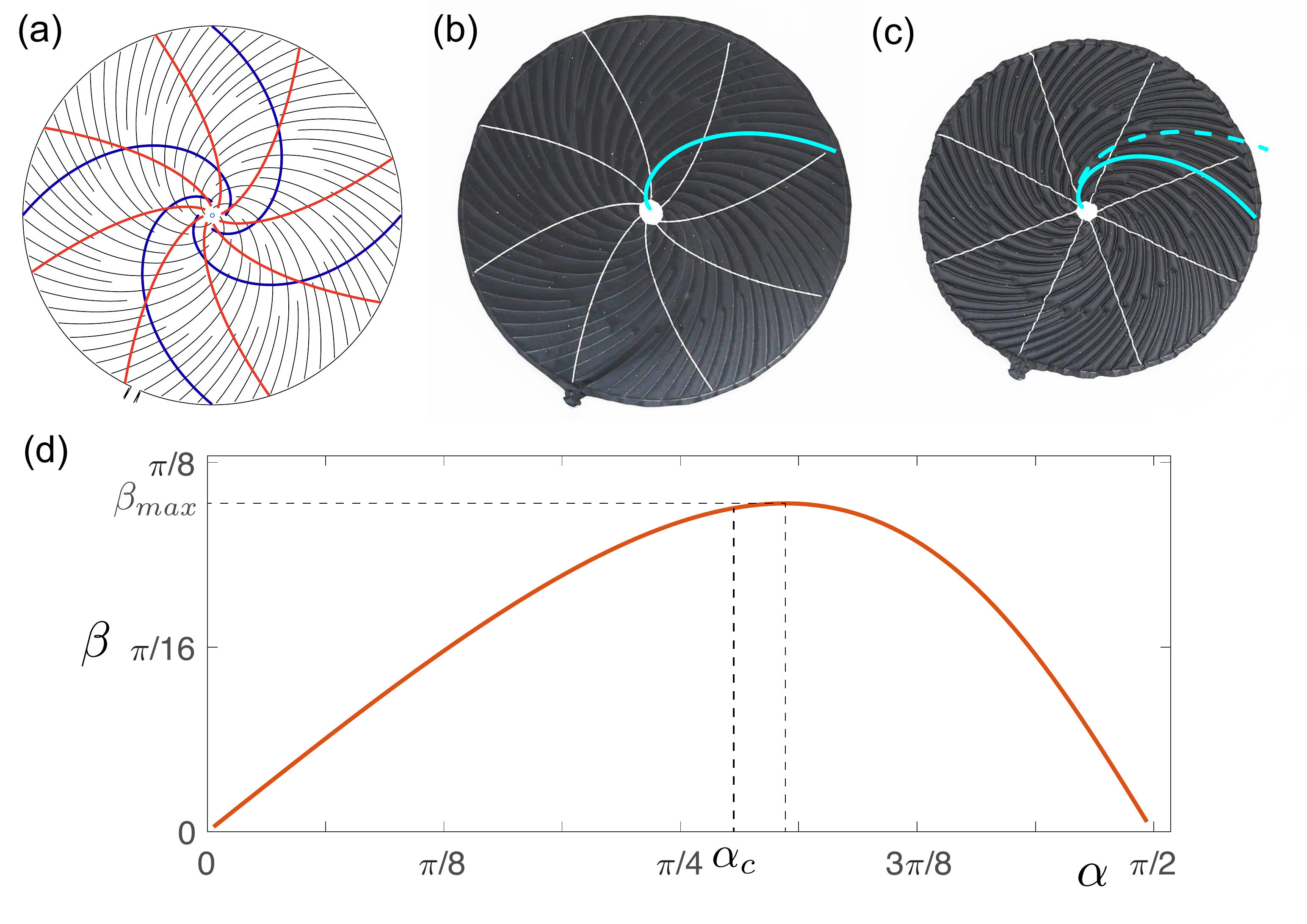}
  \caption{Evolution of proto-radii. (a) Seam pattern (black), director (blue) and proto-radii (red) of a log spiral pattern at the critical angle $\alpha\s{c}$ and a photograph (b) of the actual object at rest, with proto-radii printed in white. Upon inflation (c), the structure remains planar,  air channels evolve (blue-dotted to blue curves) and the proto-radii deform into radii (See Supplementary Video flat\_protoradii.MP4). (d) Angle $\beta$ that the proto-radii initial spirals make with the radial direction, as a function of the director angle $\alpha$ for an experimentally realistic contraction $\lambda=0.7$ (Eqn~\ref{eq:proto_log_spiral}).}
  \label{fig:proto}
\end{figure}

The cone semi-angle $\phi$ is given as before by the ratio of the new circumference $l\s{c}$, divided by $2\pi$, to the length of the new (geodesic, in-material) radius $u$. The former is
\be l\s{c}/2\pi = \sqrt{\t_c\cdot \g \cdot \t_c} =  \sqrt{ g_{\theta\theta}}\label{eq:space-radius}
\ee
 and the new radius'  length is
\be
u = \int_0^s \d s' \sqrt{\t\s{p}\cdot \g \cdot \t\s{p}} =  \int_0^r \d r' \sqrt{g_{rr} + 2 \frac{\d\theta}{\d r}  g_{\theta r}  + \left(\frac{\d\theta}{\d r}\right)^2 g_{\theta\theta} }
\ee
where $\t\s{p}$ is the tangent vector of the proto-radius, and we have taken out a $\d r/\d s$ to change $\int\d s$ to $\int\d r$. Putting in the first part of eqn.~(\ref{eq:geodesic}), $\d\theta/\d r = -g_{\theta\theta}/g_{\theta r} $, along the proto-radius, we obtain:
\be
u =  \int_0^r \d r' \sqrt{\det{\g}}/\sqrt{g_{\theta\theta} }= \lambda  \int_0^r \d r'  \frac{r'}{\sqrt{g_{\theta\theta} }} \left[\equiv   \frac{\lambda r}{\sqrt{g_{\theta\theta}/r^2 }}\right]. \label{eq:geodesic-radius}
\ee
We have used an important invariant, $\det{\g} = \lambda^2 r^2$ in polar coordinates that expresses the areal change for this 2-D metric tensor. The final equality on the RHS of eqn.~(\ref{eq:geodesic-radius}) with  $\left[ \dots \right]$ only holds for cones, where $\sqrt{g_{\theta\theta}/r^2 } = \textrm{const.}(\lambda,\alpha)$. The cone semi- angle is then given by
\be
\sin\phi = l\s{c}/(2\pi u) = g_{\theta\theta}/(\lambda r^2) = \frac{1}{\lambda} (\cos^2\alpha + \lambda^2\sin^2\alpha) = \frac{1}{\lambda} (\cos^2\alpha + \lambda^2\sin^2\alpha) \label{eq:cone_semi}.
\ee
Log spiral channels yield flat sheets at an $\alpha\s{c} = \sin^{-1}\left(\frac{1}{\sqrt{1+\lambda}}\right)$ that divides the response between cones and anti-cones.
See figure~\ref{fig:log-spirals} for a selection of channel/director angles and varying response.
Experimentally, we observe a $\lambda = 0.70$, that is about 10\% greater than $2/\pi$, which shows the presence of seams and bend etc.; see fig.~\ref{fig:set_up}.  For such a $\lambda$ the critical angle is expected to be $\alpha_c =50^\circ$. We indeed observe that the structure stays flat for such an angle. The supplementary material video flat-log-spiral.MP4 shows the critical system remaining flat under deformation, but displaying pronounced rotations at the same time, as the spiral channels evolve into modified log spiral channels.
This rotation phenomenon makes contact with the evolution of the proto-radius log spiral into a geodesic radius, as predicted above. In the flat state, the proto-radius appropriate to the $\alpha$ and $\lambda$ of the system was marked in white paint, which can be seen in figure~\ref{fig:proto}, whereupon it evolved precisely to a radial geodesic on inflation. See also the supplementary material film flat\_protoradii.MP4.

Such log spiral directors  have been imprinted in LC elastomers and glasses to produce cones and anti-cones, most notably by the Broer and by the White groups; see the review \cite{White-Broer:15}. They have been produced in arrays to give super-strong actuators that can lift several thousand times their own weight \cite{guin2018_lifters} because stretch rather than bend predominates when evolution to a new metric is frustrated by a load.

\subsection{The inverse problem -- from desired shape to required channel spirals.}\label{subsect:inverse}
The inverse problem is in general more difficult than the forward problem. A straightforward set of surfaces are those with constant Gaussian curvature, $K$, since one knows in advance what the shapes are -- spherical caps and spherical spindles for positive constant GC, and pseudo- or hyperbolic spheres for the case of axially-symmetric constant negative GC. Curvature arises from the spatial variation of the metric tensor, which we take to be in polars and where the only variation is via the angle $\alpha(r)$. In that event \cite{Mostajeran2015,Mostajeran2016} the GC is given by
\begin{equation} \label{eq:Gauss}
K
=\frac{\lambda^{-2}-\lambda^{2\nu}}{2}\left[\left(\alpha''+\frac{3}{r}\alpha'\right)\sin
(2\alpha)+2\alpha'^2\cos (2\alpha)\right].
\end{equation}
For constant $K$, eqn.~(\ref{eq:Gauss}) is an ODE for $\alpha(r)$ with a simple solution (see \cite{Mostajeran2016} for full details) :
\begin{equation} \label{eq:solution}
\alpha(r)=\pm\frac{1}{2}\arccos\left(-C(K)\,\frac{r^2}{2}+c_1+\frac{c_2}{r^2}\right),
\end{equation}
where
$C(K)=K/(\lambda^{-2}-\lambda^{2\nu})$,
and $c_1$, $c_2$ are real constants of integration. If $c_2 = 0$ one has constant Gaussian curvature on discs, otherwise for $c_2 \ne 0$ one has annular domains for which there are rich possibilities that we address elsewhere. For  discs, a real $\alpha $ requires $-1 \le c_1 \le 1$, and if  $K>0$, then
solution (\ref{eq:solution}) defines a director pattern on the  disc $r \leq \sqrt{\frac{2(1+c_1)}{C(K)}}$.  [Analogous shape evolution, but starting rather from cylinders instead of flat sheets, and with entirely different transformation mechanisms and characteristics, are addressed by Arroyo and DeSimone \cite{Arroyo2014}.]

\begin{figure}
\centering
\includegraphics[width=\linewidth]{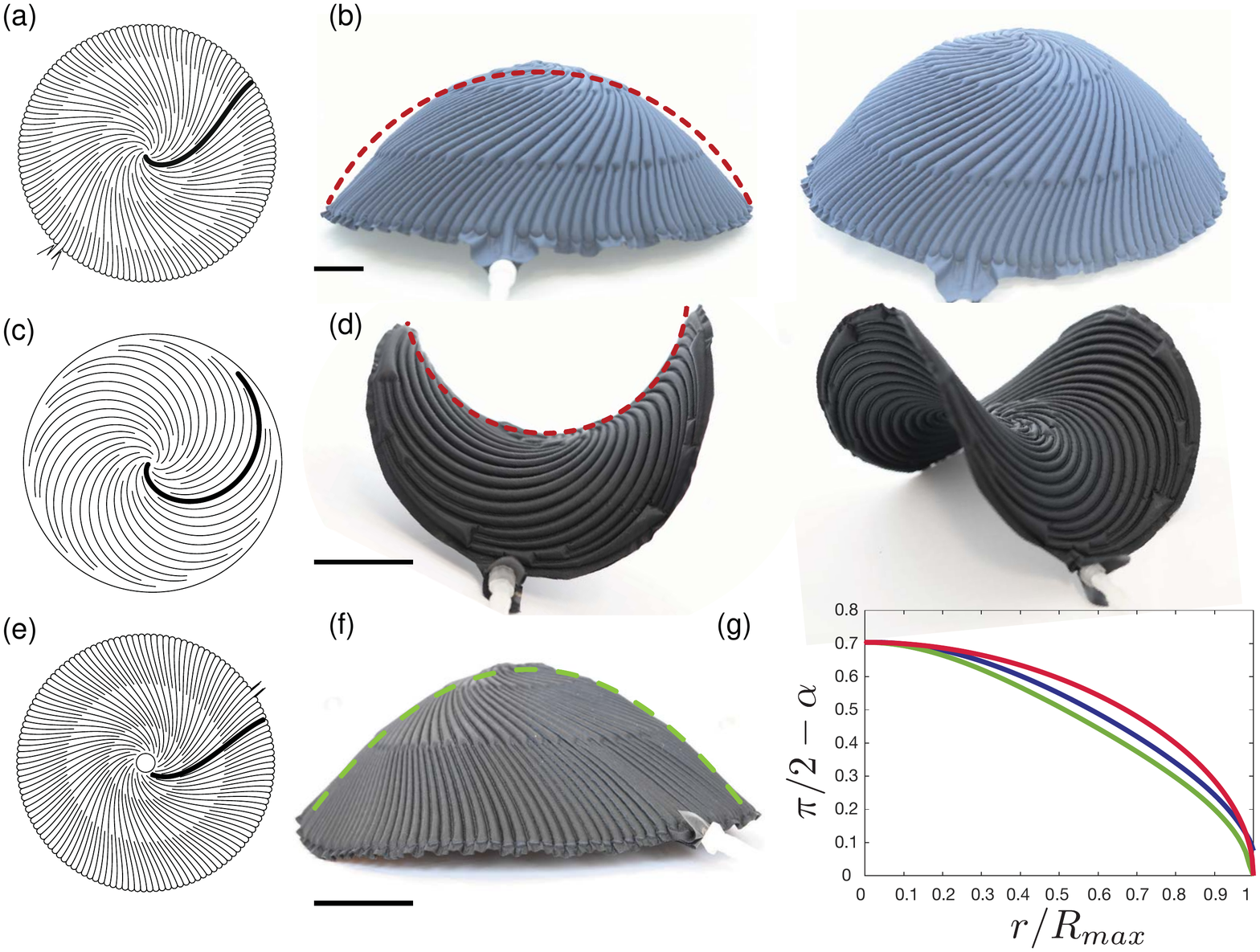}
  \caption{Inverse problems with spirals (highlighted are the channels, the directors being the duals). (a) Channel pattern for a target dome of constant positive GC, with $c_1= -(1-\lambda)/(1+\lambda)=-0.16$ (b) Photographs of the corresponding inflated structure compared with the target profile (dashed line).    (See Supplementary Video dome\_inflation.MP4)  (c \& d) Equivalent patterns and realisations for negative GC.  (e) Channel pattern for a paraboloid. (f) Corresponding inflated structure compared with the target profile (dashed line). (g) Angle of the seam lines as a function of the normalized  radial distance $r/R_{max}$ for various target shapes: (red = portion of a sphere (a)-(b), blue = catenoid and green =  paraboloid (e)-(f)).  Structures are made of TPU-impregnated nylon fabrics from Extremtextil. Scale bars: 5 cm.}
  \label{fig:axisym}
\end{figure}

To realise  spherical spindles, one has to restrict to $-1 \le c_1\leq -(1-\lambda)/(1+\lambda)$, with the upper equality giving a spherical cap where the director angle at the origin is $\sin(\alpha(r=0))=1/\sqrt{1+\lambda}$; see fig.~\ref{fig:axisym}.
There is a maximum possible reference disc radius for a given $\lambda$ that transforms into a spindle: The maximal circumference contracts by a factor of $\lambda$ (since the director there has become circumferential) to become a line of latitude on the sphere. These get relatively shorter compared with the in-material radius (think of those near the South pole which contract to zero!), and the contraction  $\lambda = 2/\pi$ cannot suffice.

If $K<0$ and $c_2=0$ one develops pseudo-spheres, that is, surfaces of constant negative GC. The bounds on $c_1$ are as above, with $c_1=-1$  yielding the spiral pattern and pseudo-sphere of maximal radius.

\subsection{The general, axi-symmetric inverse problem.} \label{subsect:general}
The general inverse problem without axial symmetry is extremely difficult but has been attacked in the arena of LC solid sheets \cite{aharoni2014geometry,aharoni2018,Griniasty2019}, with computational schemes developed to deliver a required director field for a target shape. Further, Griniasty \textit{et al} \cite{Griniasty2019} prove the existence of a local solution to the inverse problem (and provide algorithms to find all smooth director fields that deform into a desired surface geometry). In this context of pneumatic channel systems, the problem has also been attacked computationally \cite{Siefert_backpacks2019}.

Inverse problems for LC sheets with axial symmetry are less difficult and were first attacked by Aharoni \textit{et al} \cite{aharoni2014geometry} who give examples based on cartesian director patterns. In \cite{Mostajeran2018} a scheme is given  for such shapes which, in general, involves non-linear integral equations. For shells where one can express the in-material (geodesic) radius $u$ in terms of the in-space radius $\gamma_1$, that is $u = f(\gamma_1)$, simple ODEs arise. Griniasty \textit{et al} \cite{Griniasty2019} also provide the director pattern for various illustrative axisymmetric shapes.

\begin{figure}
\centering
\includegraphics[width=.35\linewidth]{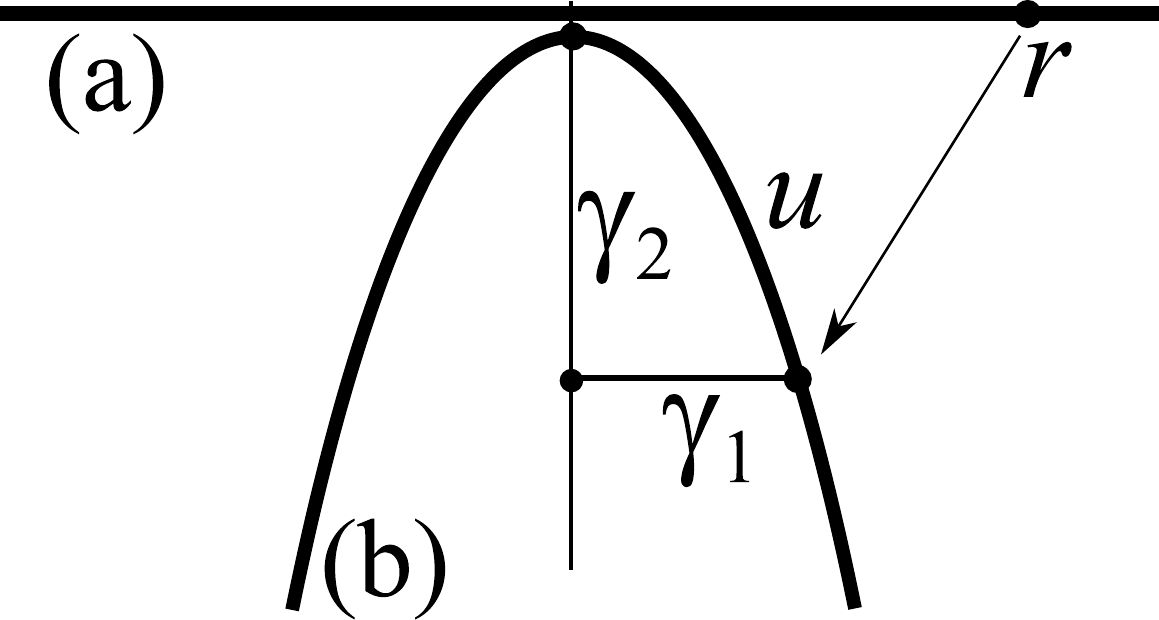}
  \caption{ A reference disc (a) where a point at radius $r$ maps, on inflation, to a point on a shell (b) given by $(\gamma_1, \gamma_2)$ and geodesic radius $u$.  }
  \label{fig:shells}
\end{figure}

Here we follow the ideas of section~\ref{subsect:log-spirals} to calculate lengths of geodesic radii in evolving structures, and the associated circumferential inflation, that is, the approach of \cite{Mostajeran2018}. We illustrate the simple ODE case with the example of a paraboloid and give an explicit $\alpha(r)$ pattern (Catenoids succumb to a similar technique.):
\nl
Consider figure~\ref{fig:shells}(a) where the radial point $r$ maps to $u$.
 Equation~(\ref{eq:geodesic-radius}) connects $r$ and $u$, and we have $\d u/\d r = \lambda r /\sqrt{\gth(r)}$.  The new circumference divided by $2\pi$ was given by  Eqn.~(\ref{eq:space-radius}), that is $\gamma_1 = \sqrt{\gth}$. In the curve of revolution we have $(\d u)^2 = (\d \gamma_1)^2 + (\d \gamma_2)^2$. Differentiating with respect to $r$, and using  $\d u/\d r$ above, we obtain:
 \bea
  \lambda r /\sqrt{\gth(r)}&=& \d\gamma_1/\d r \sqrt{1+ (\d\gamma_2/\d\gamma_1)^2} =\half \frac{\d\gth/\d r}{\sqrt{\gth}} \sqrt{1+ (\d\gamma_2/\d\gamma_1)^2} \nonumber \\
\rightarrow  \lambda r &=& \half \d\gth/\d r \sqrt{1+ (\d\gamma_2/\d\gamma_1)^2}\label{eq:gen-ode}
 \eea
which can be useful for a wide range of curves for which one knows $\d\gamma_2/\d\gamma_1$.

For paraboloids obtaining from the revolution of $\gamma_2 = \half  a \gamma_1^2$, we have $(\d\gamma_2/\d\gamma_1)^2 = a^2\gamma_1^2 = a^2 g_{\theta\theta}$, whereupon one can integrate Eqn.~(\ref{eq:gen-ode}) with respect to $r$ and $\gth$:
\be
\lambda r^2 = \frac{2}{3 a^2}\left[ (1 + a^2 \gth)^{3/2} - 1\right]. \label{eq:theta-metric}
\ee
Recall that $\gth = r^2 (1 -(1-\lambda^2)\sin^2\alpha)$ (and indeed $\gth(r=0) = 0$, which is needed for the above integration). Equation~(\ref{eq:theta-metric}) can then be solved for $\alpha(r)$ in the form:
\be
\sin^2\alpha_{\lambda}(r) = \frac{1}{1-\lambda^2} \left( 1 - \frac{1}{a^2 r^2}\left[ (1 + \frac{3 \lambda}{2} a^2 r^2)^{2/3} - 1 \right] \right) \label{eq:alpha-paraboloid}.
\ee
The subscript appearing in $\alpha_{\lambda}(r)$ is to remind one that the form of $\alpha(r)$ only gives a paraboloid for this particular target $\lambda$.
See fig.~\ref{fig:axisym} for a channel pattern for a paraboloid and the resulting inflated structure, along with the seam line angle plotted against reduced radius for a given $\lambda$.

Inverted catenoids are the ideal, self-supporting shells. We consider those that are the revolution about the $\gamma_2$ axis of the catenary $u = \frac{1}{a}\sinh(a\gamma_1)$ and where $\gamma_2 = \frac{1}{a}\cosh(a\gamma_1)$. Returning to the relation $\d u/\d r = \lambda r /\sqrt{\gth(r)} = \frac{1}{\d r} \frac{1}{a}\cosh(a\gamma_1)$, and remaining this time with the variable $\gamma_1 \equiv \sqrt{\gth}$, one has the equations:
\bea
\lambda r \d r &=& \d \gamma_1 \gamma_1 \cosh(a \gamma_1) \nonumber \\
\rightarrow \half \lambda r^2 &=& \frac{\gamma_1}{a}\sinh(a\gamma_1) - \frac{1}{a^2} \cosh(a\gamma_1) + \frac{1}{a^2} \label{eq:catenoid-implicit}.
\eea
Since  $ \sqrt{\gth} = r\sqrt{1 - (1-\lambda^2)\sin^2\alpha(r)}$, the above is an implicit equation for $\alpha_{\lambda}(r)$. Taking the limit $r \rightarrow 0$, one obtains the minimum $\alpha$ as $\sin^2\left[\alpha_{\lambda}(0)\right] = 1/(1 + \lambda)$. At the extremity of the disc that can be inflated to a catenoid, the director is tangential and $\sin^2\alpha_{\lambda}(r\s{max}) = 1$, whence $\gamma_{1 \textrm{max}} = \lambda r\s{max}$ which should be inserted into Eqn.~(\ref{eq:catenoid-implicit}) to give an equation for $r\s{max}$.   Fig.~\ref{fig:axisym}(g) shows the catenoid seam line angle against reduced radius for a given $\lambda$.

\section{Summary, conclusions}

We introduce welded systems of channels in flat, light, air-tight fabric which on inflation produce topography that has either localised or continuously distributed Gaussian Curvature (GC). Discrete variation of unform direction regions, on crossing boundaries, produces facetted structures, much like origami structures. But the GC at vertices (angular deficits) cannot be undone by unbending hinge-like folds as in conventional origami, since the origin of such deficits is the non-isometric transformation associated with inflation. Continuous variation of channel direction, here as spirals, gives distributed GC and also a variation of the direction of welds, inhibiting bend at welds (which cannot be achieved along lines that also curve). There are thus two factors contributing to increased strength.

Our experiments confirm the role of deformation in setting the limits to angular deficits and thus curvature, either around vertices or resulting from continuous field variation. Thus cubes cannot be fully achieved with a $\lambda$ geometrically bound by $2/\pi$, and nor can tetrahedra, but other solids can be. We suggest the first steps to more arbitrary facetted shells by prescribing the conditions on irregular vertices. For continuous channel variation, our achieved shells indeed have Gaussian Curvature even though their initial states are flat. The evolution of curves from and to geodesics in these curved surfaces has been experimentally observed, confirming this picture of metric-driven mechanics.

Future directions include non-simply connected shells, where new possibilities arise, and non-isometric kirigami. Initial experiments are encouraging and will be reported elsewhere. We are also concerned with the breaking of up-down symmetry where complex systems can be impeded in their path to the desired final state. The combination of differing fields along curved interfaces also offers entirely new possibilities. Large systems, several metres across, are also achievable.

\begin{ack}
MW acknowledges many discussions about non-isometric origami with Carl Modes, and about deforming spiral systems with Cyrus Mostajeran.
ES acknowledges many discussions about metric distortion and inflatables with Beno\^{i}t Roman and Jos\'{e} Bico.
\end{ack}
\begin{ethics}
No human or biological systems were used in this work.
\end{ethics}
\begin{dataccess}
Videos referred to in the text of systems transforming under inflation are available as electronic Supplementary Material at 
\nl \url{www.tcm.phy.cam.ac.uk/~mw141/eggbox\_top\_view.MP4}
\nl \url{www.tcm.phy.cam.ac.uk/~mw141/flat\_protoradii.MP4}
\nl \url{www.tcm.phy.cam.ac.uk/~mw141/flat-log-spiral.MP4}
\nl \url{www.tcm.phy.cam.ac.uk/~mw141/dome\_inflation.MP4}
\end{dataccess}
\begin{competing}
We have no competing interests.
\end{competing}
\begin{funding}
MW's work was supported by the EPSRC [grant number    EP/P034616/1].
\end{funding}
\begin{aucontribute}
ES did  both theory and experiment,  MW was only involved in theoretical aspects.
\end{aucontribute}
\bibliographystyle{rspa}

\providecommand{\noopsort}[1]{}\providecommand{\singleletter}[1]{#1}%

\end{document}